\documentclass[
amsmath,amssymb,
twocolumn,
aps,
]{revtex4-2}

\usepackage{graphicx}
\usepackage{dcolumn}
\usepackage{bm}
\usepackage[utf8]{inputenc}
\usepackage{amsmath}
\usepackage{amsfonts}
\usepackage{appendix}
\usepackage{stmaryrd}
\usepackage{mathtools}
\usepackage{comment}
\usepackage{xcolor}
\usepackage{subcaption}
\usepackage{multirow}
\usepackage{booktabs}
\usepackage[margin=0.8in]{geometry}
\usepackage{amsthm}
\usepackage{float}
\definecolor{darkgreen}{rgb}{0.0, 0.6, 0.0}

\DeclarePairedDelimiter\ceil{\lceil}{\rceil}
\DeclarePairedDelimiter\floor{\lfloor}{\rfloor}
\newtheorem{theorem}{Theorem}
\newtheorem{property}{Property}
\newtheorem{lemma}{Lemma}

\begin{document}

\preprint{APS/123-QED}

\title{ New designs of linear optical interferometers with minimal depth and component count }
\author{Timothée Goubault de Brugière}
\affiliation{Quandela, 7 Rue Léonard de Vinci, 91300 Massy, France}
\email{timothee.goubault@quandela.com}
\author{Rawad Mezher}
\affiliation{Quandela, 7 Rue Léonard de Vinci, 91300 Massy, France}
\author{Sebastian Currie}
\affiliation{Quandela, 7 Rue Léonard de Vinci, 91300 Massy, France}
\author{Shane Mansfield}
\affiliation{Quandela, 7 Rue Léonard de Vinci, 91300 Massy, France}

\date{\today}

\begin{abstract}
    We adapt an algorithm for CNOT circuits synthesis based on the Bruhat decomposition to the design of linear optical circuits with Mach-Zehnder interferometers (MZI). The synthesis algorithm reduces to designing sorting networks with nearest neighbor swapping operations as elementary gates. We recover previous designs from the literature but with additional theoretical properties regarding the compiler that implements unitaries on the interferometer. Notably the compiler can always decide whether a unitary can be implemented on a given interferometer and, if so, returns the shallowest possible implementation. We also show natural extensions of our framework for boson sampling experiments and for the coupling of multiple integrated interferometers to design larger linear optical systems. In both cases, the designs are optimal in terms of number of optical components. Finally, we propose a greedy design which exploits the arbritrary-but-fixed coupling of separate integrated interferometers to perform shallow boson sampling. We discuss the optimal interferometer dimensions to maximize the transmission. Beyond boson sampling, our developed framework allows a resource-favourable implemention of \emph{any} non-adaptive linear optical quantum algorithm, by providing the shallowest possible  interferometer for implementing this algorithm.
    
\end{abstract}

\maketitle

\section{Introduction}

Integrated photonics is a promising hardware candidate for both noisy-intermediate scale (NISQ) tasks like boson sampling \cite{aaronson2011computational}, and universal fault-tolerant quantum computing \cite{knill2001scheme, bartolucci2023fusion}. In NISQ photonics architectures computation is typically performed by passing photons through a lithographically defined integrated interferometer and detecting the photons at the output. Recent improvements in state of the art photon sources \cite{somaschi2016near} and integrated interferometers \cite{taballione2021universal} have resulted in demonstrations of photonic NISQ calculations  \cite{fyrillas2023certified,maring2024versatile}. \\

The design of these integrated interferometers is fixed upon manufacture, and control of the implemented operations is achieved by tunable phase-shifters. When designing these photonic circuits there are two key desiderata. First, we wish to ensure they are computationally universal, in the sense that they can implement any linear optical unitary \cite{reck1994experimental}, or can implement a predetermined subset of unitaries. Secondly, these circuits should minimise  photon loss and gate infidelity.  To mitigate photon loss, one minimises the optical depth of the chip. To mitigate power consumption, control complexity and fidelity errors, which arise due to noise and cross-talk in phase settings, one minimises the total number of phase-shifters. \\

For a universal $m$-mode chip with perfect beamsplitters Clements \textit{et al.} gave an optimal scheme \cite{clements2016optimal} for minimisation of both depth and number of phase-shifters. However this decomposition, although optimal, is also a normal form in that the number of components used is the same regardless of the unitary being implemented and the target application. Some specific cases are subject to improvements. For example when the beamsplitters are not perfect \cite{fldzhyan2020optimal}, when the universality constraint is relaxed, or when we can add layers of arbitrary connectivity by coupling multiple chips together.
These more general cases motivate the investigation of new circuit decompositions. \\

The design of linear optical interferometers shares a number of  similarities with the synthesis of CNOT circuits \cite{de2025shallower,DBLP:journals/cjtcs/KutinMS07, de2021gaussian}. Recent insights on the CNOT synthesis problem for a linear nearest neighbour (LNN) architecture \cite{de2025shallower} suggest a way to exploit these connections to find novel designs.  In this work we adapt the algorithms of \cite{de2025shallower} and \cite{de2021gaussian} to the case of linear optical circuit design. This results in two versatile frameworks, one based on the Bruhat decomposition \cite{borel2012linear} and one based on a greedy Gaussian elimination process. These have various applications and optimality results:
\begin{itemize}
    \item For the design of universal interferometers we recover the optimal scheme of Clements \textit{et al.} \cite{clements2016optimal}.
    \item The framework's efficient compiler decides if a target unitary can be implemented on a given LNN interferometer. The procedure always gives the shallowest implementation within the chip. 
    \item For the design of interferometers targeting Boson sampling, our framework is optimal in the number of linear optical components and in depth.
    \item For the design of universal interferometers with layers of arbritrary connectivity, our framework is optimal in the number of linear optical components.
    \item For the design of interferometers targeting Boson sampling, with layers of arbritrary connectivity, our second framework offers practical depth reduction compared to the state of the art. We also show how this practical saving in depth is useful to reach quantum utility.
\end{itemize}  

\bigskip

One interesting application of our framework concerns Boson sampling.
In particular, we show that, if we allow layers of MZIs with arbitrary connectivity, Boson sampling experiments with $m$
modes and $n$ photons can be performed using  an interferometer $V_{opt}$ of depth $O(n+\log(m))$ in worst-case. This is a significant reduction in depth as compared to, for example, a Clements interferometer $V$ performing a Boson sampling experiment, where the depth is $O(m)$. Furthermore, even when considering losses due to couplings which are present in interferometers with layers of arbitrary connectivity, we show that for a number of modes $m >> n$, the overall losses in $V_{opt}$ are still significantly smaller than in $V$. Our results therefore bring closer the possibility of experimentally achieving quantum advantage for the Boson sampling task with single photon inputs, a feat that to our knowledge has never been performed. \newline

The paper is structured as follows. Section~\ref{sec::back} gives background on linear optical circuit synthesis prior to presenting our framework based on the Bruhat decomposition in Section~\ref{sec::bruhat}. Then we discuss the compiler associated to our framework in Section~\ref{sec::compiler}. In Section~\ref{sec::extensions} we  discuss two explicit examples of practical applications of this result. In Section~\ref{sec::shallow} we present a design based on a greedy Gaussian elimination process to perform shallow Boson sampling by coupling chips together. We discuss the optimal chip size and how far we are from quantum utility in Section~\ref{sec::numerical}, before concluding in Section~\ref{sec::conclu}.

\section{Background on linear optical circuits} \label{sec::back}

Here we briefly present the linear optical circuit model before discussing known methods for implementing given unitary operations in this picture. 

\subsection{Linear optical circuits}

An interferometer on $m$ modes acts linearly on the creation and annihilation operators $a_i^{\dag}, a_i$ of each mode $i=1\dots m$. Its action is represented by a unitary matrix $U$ such that

\[ a_i^{\dag} \to \sum_{j=1}^m U_{ji} a_j^{\dag}. \] 

In other words the column $k$ of $U$ stores the image of the creation operator $a^{\dag}_k$. To implement a desired unitary $U$ on-chip, we rely on two elementary linear optical components. The beamsplitter (BS), a constant $2$-mode transformation with matrix representation 
    \[ \frac{1}{\sqrt{2}} \begin{bmatrix} 1 & i \\ i & 1 \end{bmatrix}, \]
    and the phase-shifter (PS), a parameterized $1$-mode transformation with matrix representation 
    \[ \begin{bmatrix} e^{i\phi} \end{bmatrix}. \]

We arrange these components in linear optical circuits to create more complex interferometers. The sequential composition of two operators $A$, $B$, both acting on the same subset of modes, gives the operator $BA$. The spatial composition of two operators $A$, $B$ acting on two disjoint subset of modes gives the operator $A \oplus B$ where $\oplus$ is the direct sum operator. \\

\begin{figure}
    \centering
    \includegraphics[scale=0.45]{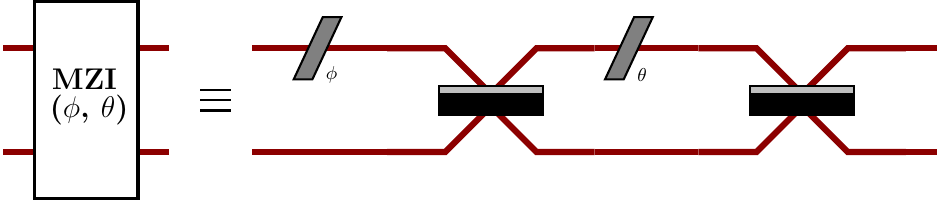}
    \caption{Implementation of an MZI in terms of phase-shifters and beamsplitters.}
    \label{fig::mzi}
\end{figure}

A beampslitter and a phase-shifter can be used to build a canonical Mach-Zehnder Interferometer (MZI). Here we define our MZI with an extra phase-shifter on the input, as shown in Fig.~\ref{fig::mzi}. This MZI is parameterized by $\phi, \theta$, and they verify the following property:

\begin{property}[\textbf{zeroing an arbitrary entry}]
Given any complex vector $\begin{bmatrix} a \\ b \end{bmatrix}$, there exists two angles $\phi, \theta$ such that 
\[ MZI(\theta, \phi) \begin{bmatrix} a \\ b \end{bmatrix} = \begin{bmatrix} a' \\ 0 \end{bmatrix}.  \]
\label{prop1}
\end{property}

For the rest of this paper, the circuits are MZI-based. Note that our results also extend to symmetric MZIs where both phase shifters are internal. This can be achieved by applying specific commutation rules to the original circuit, as done in \cite{bell2021further}.

The number of components in a circuit is the number of MZIs, and the MZI-depth of a circuit is the minimum number of time steps needed to execute a circuit, provided that two non-overlapping MZIs can be executed simultaneously. In our linear optical circuit drawings, any 2-mode box will represent the MZI in Fig.~\ref{fig::mzi}.\newline

The structure of integrated interferometers is fixed upon manufacture, and the only tuneable parameters are the angles of the MZIs. The compilation of a unitary to the interferometer therefore consists of computing angles which implement the target unitary. An $m$-mode interferometer is said to be \textit{universal} if any unitary in $\mathcal{U}(m)$, the set of unitaries of dimension $m$, can be implemented. \\

Since the structure is fixed, an integrated universal $m$-mode interferometer can only be optimal in the worst-case scenario. Whilst some unitaries in $\mathcal{U}(m)$ can be implemented with a smaller interferometer, some can't. Still, we can extend the notion of optimality to the compilation process when one wants to compute the angles required to implement a given unitary. A depth-optimal compiler would return the shallowest implementation \textit{within} the interferometer, with the following layers set to the identity. Whilst these layers will nevertheless be executed on the chip, they can be useful for mitigating errors \cite{mills2024mitigating}.\\

\subsection{Boson Sampling}
Boson sampling is the task of sampling from the output distribution of an $m$ mode interferometer implementing a Haar random unitary with $n$ input photons \cite{aaronson2011computational}. There is strong evidence that performing boson sampling on $n > 50$ photons with $n << m$ is sufficient to reach quantum utility \cite{brod2019photonic}.
Without loss of generality we can assume that our $n$ input photons enter the first $n$ modes. To implement boson sampling, the interferometer only needs to implement the images of the first $n$ creation operators. In other words, given those images in a $m \times n$ matrix $V$, any unitary $U \in \mathcal{U}(m)$ implementable by the interferometer such that 
\begin{equation} U \begin{pmatrix} I_n \\ 0_{m-n,n} \end{pmatrix} = V \label{bs_iso}
\end{equation}

is a valid implementation of the desired boson sampling experiment. An $m$-mode interferometer is said to be \textit{universal for $n$-photon boson sampling} if for any set of $n$ orthonormal columns $V \in \mathbb{C}^{m \times n}$ there exists an operator $U$ implementable on the chip such that Eq.~\ref{bs_iso} holds. \\

\subsection{Unitary decomposition schemes} \label{sec::soa}
Here we review previous decomposition schemes. We assume perfect beamsplitters. For works on imperfect beamsplitters, we refer the readers to, e.g., \cite{miller2015perfect,burgwal2017using,fldzhyan2020optimal,hamerly2022asymptotically,kumar2021mitigating,bandyopadhyay2021hardware}.

\subsubsection{QR-based schemes}

Schemes based on the QR decomposition \cite{golub2013matrix} work by zeroing the entries of $U$ sequentially until one is left with a triangular matrix. One MZI is sufficient to zero an arbitrary entry of $U$ based on Property~\ref{prop1}. Therefore one can construct a circuit composed of MZIs to progressively zero the entries of $U$. If the zeroing is done in a good order, in the sense that one zeroing does not break the zeroing of another entry, this performs a QR decomposition of $U$. \\

Triangular unitary matrices are diagonal, so once $U$ is triangular, a layer of phase-shifters can reduce $U$ to the identity. Taking the inverse of the operations gives a valid circuit for implementing $U$. Up to the diagonal matrix of $m$ parameters implemented by the layer of phase-shifters, a unitary matrix of size $m$ is parameterized by $m(m-1)$ real parameters. As our MZIs have two phase-shifters, $m(m-1)/2$ are necessary for universality. Regarding the depth, at most $m-1$ non-overlapping MZIs can be executed in two time steps, so the optimal MZI depth for a universal chip is $\frac{2m(m-1)}{2(m-1)} = m$. \newline

\begin{figure}
    \centering
    \includegraphics[scale=0.7]{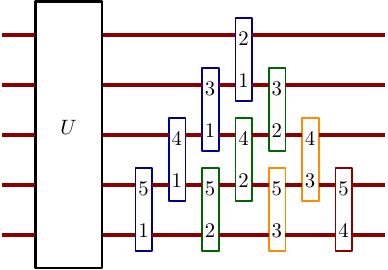}
    \caption{The process to zero an arbitrary $U$ using Reck' scheme. Each box/MZI has two numbers $i,j$ giving the entry $U_{ij}$ that is zeroed with this block. Once all MZIs have been applied, $U$ is diagonal and can be synthesized with a layer of phase-shifters.}
    \label{reck}
\end{figure}

\begin{figure*}
    \centering
    \includegraphics[scale=0.7]{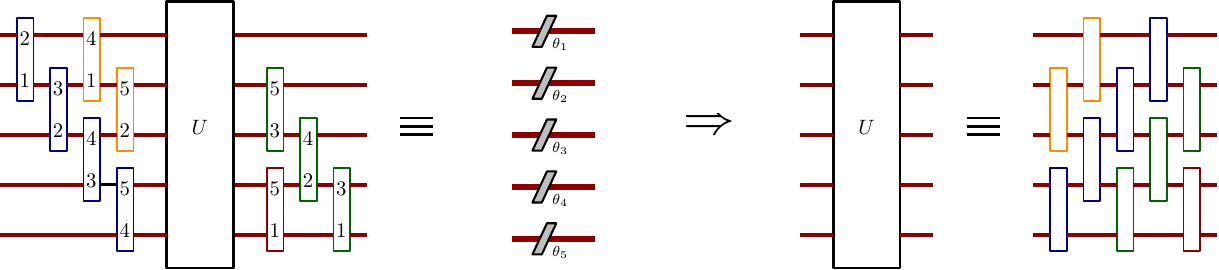}
    \caption{The process to zero an arbitrary $U$ using Clements' scheme. Each box/MZI has two numbers $i,j$ giving the entry $U_{ij}$ that is zeroed with this block. MZIs applied on the left act on the columns of $U$ while MZIs applied on the right act on the rows of $U$. Once all MZIs have been applied, $U$ is diagonal and can be synthesized with a layer of phase-shifters.}
    \label{clements}
\end{figure*}

Two main strategies for zeroing the entries of $U$ are proposed in the literature:
\begin{itemize}
    \item Reck et al.'s scheme \cite{reck1994experimental} zeroes all the elements of one column of $U$ at a time. The process is represented in Fig~\ref{reck}. This scheme requires $m(m-1)/2$ MZIs, and the depth is $2m-3$. This is optimal in number of MZIs but not in depth.
    
    \item Clements et al.'s scheme \cite{clements2016optimal} is described by the circuit in Fig.~\ref{clements}. Each diagonal layer of MZIs is used to zero one subdiagonal of $U$ and we alternate between operations on the rows and the columns of $U$ to do the zeroing in a shallow way. The layer of phase-shifters needed to implement the diagonal, initially between the left and right circuits, can be postponed to the end of the circuit using commutation rules. Inverting and merging the left and right circuits gives an circuit which is optimal in both the number of components and depth.

\end{itemize}

\ \\ 
\paragraph*{Block QR-scheme.} In \cite{su2019hybrid} a block version of Clements et al.'s scheme is proposed. This method was further improved in \cite{kumar2021unitary}. They rely on the use of suboperators in $SU(n)$ to implement larger operators in $SU(m), m > n$. Their method is generic and can handle different degrees of freedom of the light as the suboperators in $SU(n)$ can act on other modes than the spatial modes. They show that they can synthesize arbitary operators on $m$ modes by using arbitrary operators on $n$ modes and residual operators on $2n-3$ modes. 

\subsubsection{Cosine-sine decomposition based schemes}
We briefly review some works in the literature that use the Cosine-Sine decomposition \cite{golub2013matrix}. These can be useful for designing the coupling of multiple integrated interferometers. Any unitary $U$ of size $m$ can be written
\[ U = \begin{bmatrix} A_1 & \\ & A_2 \end{bmatrix} \begin{bmatrix} C & -S &  \\ S & C &  \\  &  & I_{m-2p} \end{bmatrix} \begin{bmatrix} B_1 & \\ & B_2 \end{bmatrix} \]

where $A_1, B_1, C, S$ are $p \times p, p \leq m/2$, $A_2, B_2$ are $(m-p) \times (m-p)$. $A_1, A_2, B_1, B_2$ are unitary matrices, $C, S$ are real diagonal unitary matrices. Such a decomposition exists for any $p < m$. This leads to a recursive procedure to implement $U$.

\begin{itemize} 
    \item The special case $p=1$ can be found in \cite{de2018simple}, this is the decomposition of de Guise \textit{et al.}. Their method gives results close to Reck's scheme but exhibits novel theoretical properties.
    \item Other works in the literature use the Cosine-Sine decomposition but with other block sizes \cite{dhand2015realization}. Overall, the circuit decomposes into generic operators on $k < m$ modes and special operators called CS-matrices that correspond to the central term in the Cosine-Sine decomposition. These operators are applied on $2k$ modes but an implementation with MZI remains unclear. These works do not necessarily aim at a pure spatial mode implementation but their method transposes to this case as well. These methods were improved in \cite{kumar2021unitary}.
\end{itemize}

\section{A sorting network algorithm based on the Bruhat decomposition} \label{sec::bruhat}

The core of our framework relies on the Bruhat decomposition of invertible complex matrices \cite{borel2012linear}, and how MZIs modify the decomposition. 

\begin{theorem}[\textbf{Bruhat decomposition}]
Let $A \in GL_n(\mathbb{C})$, there exists two invertible upper triangular matrices $U_1, U_2$ and a \underline{unique} permutation matrix $P$ such that 
\[ A = U_1 P U_2.  \]
Moreover, $U_1$ can be chosen such that its diagonal coefficients are all equal to $1$. 
\end{theorem}

When $A$ is unitary, and therefore invertible, the theorem holds. When $P$ is the identity matrix then $A$ is upper triangular and unitary so diagonal. This means that if we can find a circuit that implements an operator $C$ such that $CA = U_1 I U_2 = D$ then a layer of phase-shifters implementing $D$ ends the synthesis and the circuit $C^{\dag}D$ gives a valid implementation of $A$. 

Note that when inverting the circuit $C$ we end up with reversed MZIs and the layer of phase shifters implementing $D$ is at the beginning of the circuit. To recover a more standard circuit, i.e., with the usual design of MZIs with an external phase shifter on the input side and one final layer of phase shifters, one can propagate the first layer of phase shifters to the end using commutation rules. For more details see \cite[Section 4]{clements2016optimal}.
\\

We now explore how MZIs affect the Bruhat decomposition. Starting with a unitary $A$ and its Bruhat decomposition $A = U_1 P U_2$, we assume, without loss of generality, that we apply an arbitrary operator $E$ that solely acts on the first two modes of $A$. This will locally break the Bruhat decomposition of $A$, and we show how to recover the Bruhat decomposition of $EA$ with invariant local operations. Namely, let $C_1$, $C_2$ be two arbitrary invertible operators, we can always write 
\[ E A = E U_1 P U_2 = \left(E U_1 C_1\right) \left(C_1^{-1} P C_2\right)\left(C_2^{-1} U_2\right). \]

We show how to compute $C_1, C_2$ such that 
\[ E A = \left(E U_1  C_1\right) \left( C_1^{-1} P C_2 \right)  \left(C_2^{-1}  U_2 \right) = U'_1 P' U'_2 \]

is a valid Bruhat decomposition of $EA$. We can write

\[ E U_1 = E  \begin{bmatrix} 1 & a & \hdots \\ 0 & 1 & \hdots  \\ & & \ddots \end{bmatrix} = \begin{bmatrix} \alpha & a' & \hdots  \\ \beta & \gamma & \hdots \\ & & \ddots \end{bmatrix}.  \]

Assuming $E$ is not the identity operator (otherwise we would already have a valid Bruhat decomposition), we necessarily have $\beta \neq 0$. We consider two cases:

\begin{itemize}
    \item $\gamma=0$, then we simply swap the columns $1$ and $2$ of $U_1$. In other words 
    \[ C_1 = \begin{bmatrix} 0 & 1 & \\ 1 & 0 & \\ & & I_{m-2} \end{bmatrix}. \]
     Then $C_1^{-1}P$  is already a permutation matrix given from $P$ by swapping the rows $1$ and $2$. Therefore $C_2 = I$ and we do not have to modify $U_2$. \textit{Overall, we have swapped two rows of $P$}.
    \item $\gamma \neq 0$, then we need to specify again two cases. Let $j_1, j_2$ be the unique integers such that $P[1,j_1] = 1$ and $P[2, j_2] = 1$.
    \begin{itemize}
        \item[$\diamond$] $j_1 > j_2$. We zero $\beta$ with an elementary column operation 
        \[ c_1 \leftarrow c_1 - \beta/\gamma \cdot c_2 \] between columns $1$ and $2$ of $U_1$. In other words 
        \[ C_1 = \begin{bmatrix} 1 & 0 & \\ -\beta/\gamma & 1 \\ & & I_{m-2} \end{bmatrix}. \]
        We apply $C_1^{-1}$ on the rows of $P$, i.e, the elementary row operation 
        \[ r_2 \leftarrow r_2 + \beta/\gamma \cdot r_1 \] 

        which will modify the following entries of $P$ 
        \[ \begin{bmatrix} P_{1,j_1} & P_{1, j_2} \\ P_{2,j_1} & P_{2, j_2} \end{bmatrix} = \begin{bmatrix} 1 & 0 \\ 0 & 1\end{bmatrix}  \to \begin{bmatrix} 1 & 0 \\ \beta/\gamma & 1 \end{bmatrix}. \]

        To recover back a permutation matrix we do the elementary column operation 
        \[ c_{j_1} \leftarrow c_{j_1} - \beta/\gamma \cdot c_{j_2}. \] 
        Again we have to do the elementary row operation on $U_2$
        \[ r_{j_2} \leftarrow r_{j_2} + \beta/\gamma\cdot r_{j_1} \]
        which does not break the triangular structure of $U_2$ because $j_2 < j_1$ by assumption. \textit{Overall, $P$ is unchanged}.
        \item[$\diamond$] $j_1 < j_2$. We do exactly as in the case $j_1 > j_2$ except that we first swap the first two columns of $U_1$ and the first two rows of $P$ before. Then the process is the same. \textit{Overall, the first two rows of $P$ are swapped}.
    \end{itemize}
\end{itemize}

From this derivation, we highlight two key properties: 
\begin{itemize}
    \item Any operator on two neighbor modes can only locally change $P$, namely the operator can only swap the two rows of $P$ on which it is applied.
    \item One can always choose a suitable operator to swap two neighbor rows of $P$. If we want to swap rows $i$ and $i+1$ of $P$, we look at the submatrix 
    \[ \begin{bmatrix} (U_1)_{i,i} & (U_1)_{i, i+1} \\ (U_1)_{i+1,i} & (U_1)_{i+1,i+1} \end{bmatrix} = \begin{bmatrix} 1 & b \\ 0 & 1 \end{bmatrix} \] 
    and choose an MZI performing the transformation 
    \[ \begin{pmatrix} b \\ 1 \end{pmatrix} \to \begin{pmatrix} b' \\ 0 \end{pmatrix} \]
    which is always possible due to Property~\ref{prop1}. Then, referring to the case $\gamma=0$, we can swap the columns $i$ and $i+1$ of $U_1$ and swap the rows $i, i+1$ of $P$. To recover the diagonal elements of $U_1$ to $1$, we can propagate a diagonal matrix up to $U_2$.
\end{itemize}

Therefore the problem simplifies into reducing $P$ to the identity matrix with local swaps of its rows. Equivalently, if we give each row $i$ the label $j$ such that $P[i,j]=1$, we have to sort the labels using a sorting network which is LNN compliant. \\

\begin{figure}[b]
    \centering
    \includegraphics[scale=0.9]{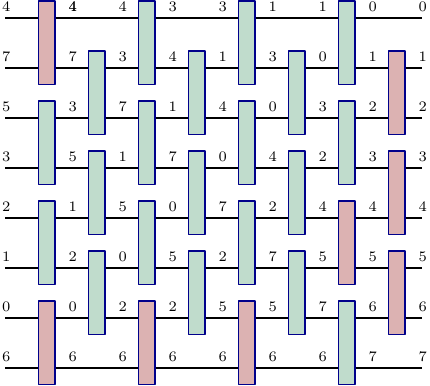}
    \caption{Sorting network for $8$ wires. Each box is a conditional swap. For this particular input, red boxes do not perform a swap and correspond to identity MZI in the resulting circuit. Green boxes apply a swap and correspond to non trivial MZI.}
    \label{fig::sorting}
\end{figure}

Depth optimal sorting networks which are LNN compliant are known and we present one to sort $8$ labels in Fig~\ref{fig::sorting}, with an example of initial label arrangements. Each box is a conditional swap. A sorting network to sort $n$ labels is of depth $n$. Translating back to the linear optical circuit model, each box is an MZI, and so the circuit is also of depth $n$. This recovers the scheme of Clements \textit{et al.} \cite{clements2016optimal}. Sorting the labels one by one recovers Reck \textit{et al.}'s design \cite{reck1994experimental}.

\section{An optimal compiler for LNN interferometers} 
\label{sec::compiler}
Given a unitary $U$ and a LNN compliant interferometer of any shape, our framework can be used to find the shallowest part of the interferometer required to implement $U$. 
This could allow one to use the remaining layers to mitigate errors \cite{mills2024mitigating}. This result is a consequence of the following theorem regarding the compiler: 
\begin{theorem}
Given a unitary $U$ and an LNN compliant interferometer that can implement $U$, there exists a procedure that can always output a set of angles for the MZIs to implement $U$.
\label{thm1}
\end{theorem}

Therefore, one can apply the procedure with different subsets of the interferometer, check if synthesis is possible, and output the shallowest circuit that works. Given an ending layer of the interferometer we will also show that the procedure will output the shallowest circuit possible ending at this layer, if synthesis is possible. This means we only need to iterate through all possible ending layers, which results in an overhead linear in the size of the interferometer.\\

To prove Theorem~\ref{thm1}, we restate it in terms of sorting networks. We focus on the labels $P$ of the modes using the Bruhat decomposition of $U$. The input interferometer is equivalent to a network of ``mixers" (one for each MZI) where each mixer can freely swap the labels of the two wires on which it is applied. If the circuit has $k$ MZIs, there are $2^k$ possible scenarios leading to a set of possible permutations $P$. We write $\mathcal{S} = \{ \sigma_K \; | \; K \in [0,1]^k \}$ the set of reachable permutations where $K_i=1$ if the $i$-th mixer swaps the labels, $K_i=0$ otherwise. \\

\begin{figure}[b]

\includegraphics[scale=0.5]{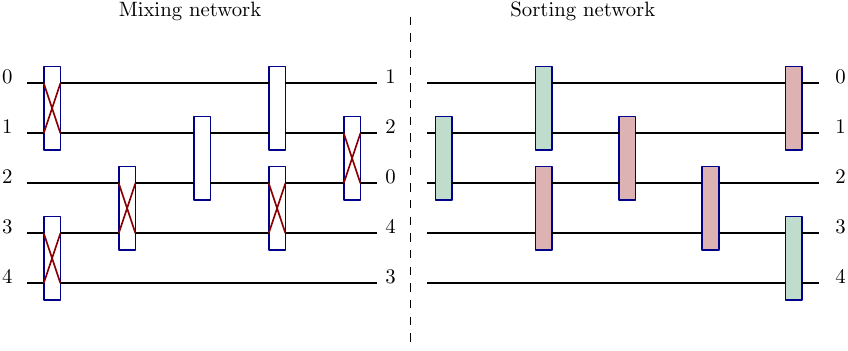}

\caption{The mixing network corresponds to an MZI circuit implementing unitary $U$. Each box potentially applies a SWAP gate on the labels, depending on the values of the angles of the MZI and on the unitary on which the MZI is applied. Shown is a possible scenario where $\sigma_{[1,1,1,0,0,1,1]} = [1,2,0,4,3]$ (reading the boxes left to right, top to bottom). In the Bruhat decomposition of $U$, we recover the labels of the wires in $P$. Then a sorting network corresponding to the reverse of the mixing network is applied. Green boxes are where SWAPs are made, red boxes are identity. Note that this is not the symmetric of the mixing network. Our result shows that for any scenario in the mixing network, the sorting network will always sort back the labels.}
\label{fig::procedure}
\end{figure}

The procedure consists in applying a sorting network given by the reverse of the MZI circuit. During this process, each MZI will apply a conditional SWAP and locally sort the labels of two wires. The process is illustrated in Fig~\ref{fig::procedure}. Theorem~\ref{thm1} is true if this procedure can always reduce $P$ to the identity for any possible $P \in \mathcal{S}$. In other words we want to prove the following theorem:

\begin{theorem}

Given a mixing network of $k$ mixers and $\mathcal{S}$ the set of reachable permutations from this network. For any $\sigma \in \mathcal{S}$, the sorting network given by the reverse of the mixing network reduces $\sigma$ to the identity permutation.

\end{theorem}

\begin{proof}

By induction on the number of mixers. The result is true for the empty mixing network. Now suppose the result is true for mixing networks with $k$ mixers. We consider a mixing network of $k+1$ mixers, and a permutation $\sigma_K = [\sigma_1, \sigma_2, ..., \sigma_{k+1}] \in \mathcal{S}$ from its set of reachable permutations. Without loss of generality, we assume that the last mixer is applied on the first two wires of the networks. Following the procedure, we apply a conditional swap on the first two wires, giving a permutation 
\[ \sigma' = [\sigma'_1, \sigma'_2, ...] \]
with 
\[ \sigma'_1 = \min(\sigma_1, \sigma_2), \sigma'_2 = \max(\sigma_1, \sigma_2). \]

With an abuse of notation we refer to $\sigma_{K[1:k]}$ to the permutation given by the first $k$ mixers of the sorting network. $\sigma'$ only differs from $\sigma_{K[1:k]}$ on the values of the first two entries, as those entries are the only ones that have been modified, first by the last mixer of the mixing network, then by the first conditional swap of the sorting network. We distinguish two cases: 
\begin{itemize}
    \item $(\sigma_{K[1:k]})_1 < (\sigma_{K[1:k]})_2$, this means that $\sigma' = \sigma_{K[1:k]}$ and $\sigma'$ is reachable from a mixing network with $k$ mixers,
    \item $(\sigma_{K[1:k]})_1 > (\sigma_{K[1:k]})_2$, this means that at some point in the mixing network there exists a mixer that swapped the labels $(\sigma_{K[1:k]})_1$ and $(\sigma_{K[1:k]})_2$. Let's assume this is the $j$-th mixer for some $j$. Inverting the operation of this mixer (therefore bitflipping $K_j$) will act as a change of variables: it will replace $(\sigma_{K[1:k]})_1$ with $(\sigma_{K[1:k]})_2$ and vice versa because the action of the mixing network does not depend on the labels of the wires. Therefore, if we write $K'$ the bitstring given from $K$ by flipping its $j$-th entry, we have $\sigma_{K'[1:k]} = \sigma'$.
\end{itemize}

In both cases, the resulting permutation $\sigma'$ belongs to the set of permutations reachable from the first $k$ mixers. By hypothesis, the sorting network given by the reverse of this mixing network will reduce $\sigma'$ to the identity. Concatenating this with the first conditional swap gives a valid sorting network that reduces $\sigma$ to the identity.

\end{proof}

The procedure sorts the labels in the shallowest way, otherwise we could find an example of interferometer that does not satisfy Theorem~\ref{thm1}. Therefore the compiler not only outputs a set of angles for implementing a given unitary on a given interferometer, it also outputs the shallowest way to do it. \\

\paragraph*{\textbf{Special case of a rectangular interferometer.}}

Our result applies to the case of a universal rectangular chip. As the shape of the chip is a succession of two alternating layers, the overhead is low as only two calls to the synthesis framework are necessary. We have the theoretical guarantee that our framework will compute the optimal depth to implement any unitary in the rectangular-shaped chip. It is not clear if the compiler given by Clements \textit{et al.}'s scheme \cite{clements2016optimal} exhibits similar property. From numerical simulations, it appears that the scheme also returns the shallowest possible implementation but to our knowledge there is not theoretical proof that it is always the case.

\section{Extensions of the framework} \label{sec::extensions}

\subsection{Application to boson sampling}

Implementing a full operator $U$ on $m$ modes is only necessary if there are input photons in every mode, which is not always the case in boson sampling experiments. If we know that the $k$-th mode will never receive a photon as input then the output statistics do not depend on the $k$-th column of $U$. We can use this to further compress the interferometer and reduce the number of active components. \newline

Let $m$ be the number of modes, and $n$ the number input photons. For simplicity we assume that the photons enter the first $n$ modes. The goal is now to implement the first $n$ columns of a full $m$-mode operator $U$. Namely we need to find an optical circuit on $m$ modes that implements an operator $C$ such that
\[ C \begin{bmatrix} I_n \\ \ \end{bmatrix} = \begin{bmatrix} u_1 & u_2 & \hdots & u_n \end{bmatrix} := V \]
where the $u_i$'s are the columns of $U$. \newline 

Each column of $V$ is parameterized by $2m$ real degrees of freedom, and in total $V$ can be parameterized by $2nm$ real numbers. However $V$ satisfies the relation $V^{\dag}V = I_n$, giving $n^2$ equations removing $n^2$ degrees of freedom. Therefore, again up to a diagonal matrix that removes $n$ parameters, we need at most 
\[ 2nm - n(n+1) \]
real numbers to characterize $V$. Hence we need only $nm - n(n+1)/2$ MZIs to implement $V$. Regarding the MZI-depth, in the worst case, as we still want the last mode to talk to the first one, the optimal depth is lower bounded by $m-1$. \newline

Current methods cannot reach both an optimal depth and an optimal number of MZIs as summarised in Table~\ref{boson_sampling_recap}. The naive way to adapt the methods to partial synthesis is to remove the MZIs that never encounter a photon. Otherwise, extending these methods to partial synthesis is not easy. For example a Clements decomposition  requires one to perform column operations that are no longer allowed. \newline 

\begin{table*}

    \begin{tabular}{cccc}
    \toprule
    Scheme & Circuit & Number of MZIs & MZI-depth \\
    \hline
    \hline
    \\
    \hspace*{0.5cm} Reck \textit{et al.} \cite{reck1994experimental} \hspace*{0.5cm} & \hspace*{0.5cm} \begin{tabular}{c} \includegraphics[scale=0.5]{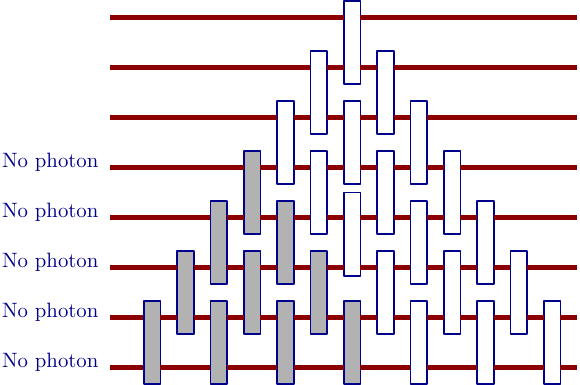} \end{tabular} \hspace*{0.5cm} & \begin{tabular}{c} $nm - \frac{n(n+1)}{2}$ \\ \\ (\textcolor{darkgreen}{optimal}) \end{tabular} & \begin{tabular}{c} $m+n-2$ \\ \\ (\textcolor{red}{non optimal}) \end{tabular} \\
    \midrule
    \hspace*{0.5cm} Clements \textit{et al.} \cite{clements2016optimal} \hspace*{0.5cm} & \hspace*{0.5cm} \begin{tabular}{c} \includegraphics[scale=0.5]{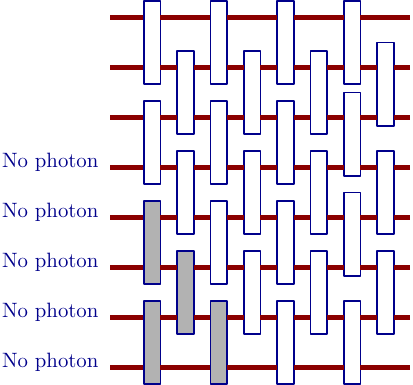} \end{tabular} \hspace*{0.5cm} & \begin{tabular}{c} $\approx \frac{m^2-n^2+2mn}{4}$ \\ \\ (\textcolor{red}{non optimal}) \end{tabular} & \begin{tabular}{c} $m$ \\ \\ (\textcolor{darkgreen}{optimal}) \end{tabular} \\
    \midrule
    \hspace*{0.5cm} Our work \hspace*{0.5cm} & \hspace*{0.5cm} \begin{tabular}{c} \includegraphics[scale=0.5]{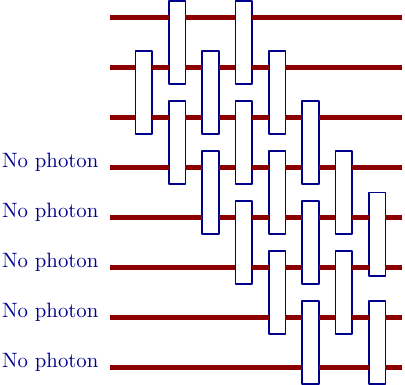} \end{tabular} \hspace*{0.5cm} & \begin{tabular}{c} $nm - \frac{n(n+1)}{2}$ \\ \\ (\textcolor{darkgreen}{optimal}) \end{tabular} & \begin{tabular}{c} $m$ \\ \\ (\textcolor{darkgreen}{optimal}) \end{tabular} \\
    \bottomrule
    \end{tabular}

    \caption{Extension of Reck's and Clements' schemes for $n$-photon $m$-mode Boson sampling and comparison with our own proposal. For Reck and Clements' circuits, the black boxes are the MZIs that are removed by eye from the original scheme as they never encounter a photon.}

    \label{boson_sampling_recap}
    
\end{table*}

We can however adapt the algorithm presented in Section~\ref{sec::bruhat} to this case, and find a circuit with same depth but with the optimal number of MZIs. If we manage to sort the first $n$ labels, then the Bruhat decomposition in a block form where we separate the first $n$ columns would give 
\[ A = \begin{pmatrix} U_1 & B \\ 0 & U_2 \end{pmatrix} \begin{pmatrix} I & 0 \\ 0 & P \end{pmatrix} \begin{pmatrix} U_3 & C \\ 0 & U_4 \end{pmatrix} = \begin{pmatrix} U_5 & D \\ 0 & E \end{pmatrix} \]
where $U_1, U_2, U_3, U_4, U_5$ are all triangular. The unitarity of $A$ implies that $D = 0$ and $U_5$ is diagonal which ends the synthesis, as we only care about reducing the first $n$ columns to the identity operator. What then remains is to provide a simplified sorting network that can always sort the first $n$ labels with fewer conditional swaps, meaning fewer MZIs in the resulting circuit. \newline

\begin{figure*}
    \centering
    \begin{subfigure}{0.32\textwidth}
        \centering
        \includegraphics[height=1.2in]{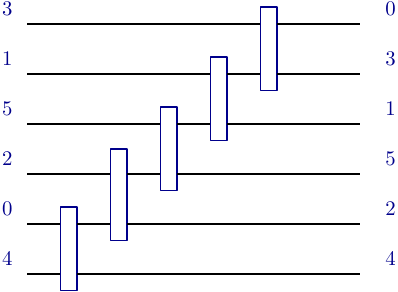}
        \caption{Partial sorting network to sort 1 label.}
        \label{fig:partial_sorting_1}
    \end{subfigure}
    \begin{subfigure}{0.32\textwidth}
        \includegraphics[height=1.2in]{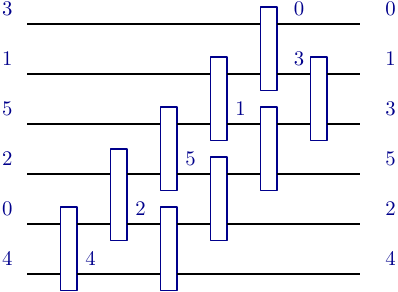}
        \caption{Partial sorting network to sort 2 labels.}
        \label{fig:partial_sorting_2}
    \end{subfigure}
    \begin{subfigure}{0.32\textwidth}
        \includegraphics[height=1.2in]{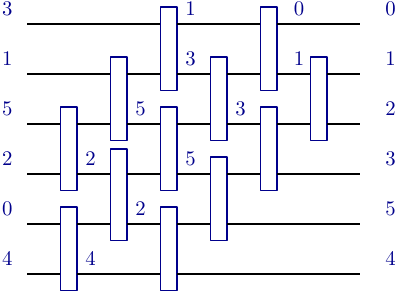}
        \caption{Partial sorting network to sort 3 labels.}
        \label{fig:partial_sorting_3}
    \end{subfigure}
    \caption{Partial sorting networks to sort the first $k$ labels only, $k=1..3$.}
    \label{fig::partial_sorting}
\end{figure*}

We give a recursive procedure to construct partial sorting networks, with an example shown in Fig.~\ref{fig::partial_sorting}. 

\begin{itemize}
    \item \textit{\underline{Initialization}:} to sort one label, the circuit given in Fig.~\ref{fig:partial_sorting_1} is optimal as the last mode needs to connect to the first one.
    
    \item \textit{\underline{Adding one label}:} adding a label to sort can be done by adding a subdiagonal or superdiagonal layer of conditional swaps. In our example, we need a subdiagonal layer to sort a second label, shown in Fig.~\ref{fig:partial_sorting_2}. If we have a circuit sorting $k$ labels, adding a subdiagonal layer is equivalent to adding the layer sequentially to the circuit. Overall, the circuit was sorting the first $k$ labels and the last layer can be seen as sorting the first label of the remaining $n-k$ labels. Thus, label $k+1$ will end in the $k+1$-th wire and will be sorted.
    
    \item \textit{\underline{Adding another label}:} If we want to sort a third label, we need to add a superdiagonal layer of conditional swaps like in Fig~\ref{fig:partial_sorting_3}. Starting from a circuit sorting $k$ labels, adding a superdiagonal layer lets us sort one label among the first $n-k$ ones. Necessarily, one of them belongs to the first $k+1$ labels. We move it to the first wire with our newly added layer of conditional swaps. Then the remaining circuit is a sorting network for $k$ labels, it will sort the remaining unsorted $k$ first labels. During this process, the originally $k+1$-th sorted label cannot be moved outside the first $k+1$ wires. Overall we have sorted the first $k+1$ labels.
\end{itemize}

\ \\
To sort $n$ labels, we need 
\[ m-1 + m-2 + \hdots + m-n = mn - \frac{n(n+1)}{2} \]
MZIs, which is optimal. The depth of the circuit is $m$, which is one unit more than the lower bound. Provided that $n << m$, which is the case in boson sampling experiments, this new scheme provides a substantial reduction in the number of linear optical components, from $O(m^2)$ to $O(mn)$, at the expense of potentially inducing unbalanced loss across the modes. 

\subsection{Coupling multiple interferometers}
Whilst integrated photonics allows for large numbers of components on-chip, this is not unbounded. There is an upper limit on the size of interferometer manufacturable, which depends on the modal confinement, the length of the phase-shifters, and the size of the chip. To reach the size of interferometers required for tasks like boson sampling, one can consider the networking of many smaller integrated interferometers. This affords us layers of arbitrary but fixed connectivity, as the fibres connections can in-principle couple any mode to any other.  This comes at the cost of needing to couple between chips, which induces loss. The fibre links will also be susceptible to phase instability.

Combining several smaller chips to the design of a larger chips is not new and relative works can be found in \cite{kumar2021unitary, dhand2015realization}. Some limitations of these works are that they either rely on special linear optical components that may not be easy to implement in practice, or they rely on other encodings (for instance polarization). \newline

\begin{figure}
    \centering
    \includegraphics[scale=0.75]{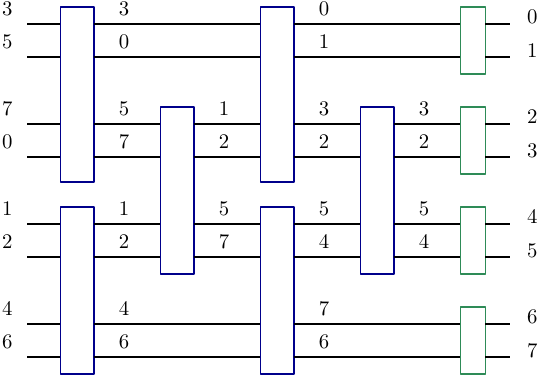}
    \caption{Block sorting network. Each blue box is a sorting network that sorts a block of labels. Green boxes are universal interferometers.}
    \label{fig::block_sorting}
\end{figure}

Using our framework we can design sorting networks that combine several smaller interferometers. To perform universal computations we propose to sort the labels by blocks. An example is given in Fig.~\ref{fig::block_sorting} with 8 modes and blocks of $2$ modes. Each blue interferometer on $4$ modes creates two sorted blocks of labels. The arrangement of the blue interferometers follows a sorting network on $4$ modes. Ultimately we need to sort the labels within each block with universal interferometers in green.

With an $m$-mode computation and $k$ blocks of size $p$ (assuming $m=kp$), the network has $k(k-1)/2$ interferometers on $2p$ modes arranged as an interferometer-circuit of depth $k$ and $k$ interferometers on $p$ modes counting as an interferometer-circuit of depth $1$. \newline

Any $2p$-mode universal interferometer can be used to sort the block of labels. This gives a global circuit with 
\[ \frac{k(k-1)}{2} \frac{2p(2p-1)}{2} + k \frac{p(p-1)}{2} \approx m^2 \] MZIs and an MZI-depth of 
\[ k \times 2p + p = 2m + p. \] 
Note that the depth and the number of components have doubled compared to the standard scheme. \\

\begin{figure}
    \centering
    \includegraphics{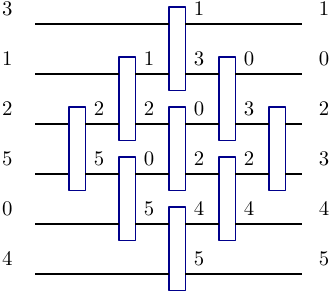}
    \caption{Diamond design on $6$ modes for block sorting the labels.}
    \label{fig::diamond}
\end{figure}

To improve the number of components, we use smaller interferometers to sort the blocks of labels. We want to separate the labels into two blocks: the ones with lowest values and the ones with highest values. This is illustrated in Fig~\ref{fig::block_sorting} by the wire labels. The labels within a block do not need sorting until the very end. We can use the diamond design shown in Fig~\ref{fig::diamond} for 6 modes to do this partial sorting. One can see this design as a sequence of diagonal layers, each one moving one label from the lower block to the upper one. This new $2p$-mode interferometer has $p^2$ MZIs and depth $2p-1$. The total number of MZIs required for universality becomes
\[ \frac{k(k-1)}{2} p^2 + k \frac{p(p-1)}{2} = \frac{m(m-1)}{2} \]
which is optimal. The MZI-depth is now
\[ \left( 2 - \frac{1}{p} \right) m + p \]
which is not asymptotically optimal unless $p=1$.

\section{A shallow gaussian elimination framework for Boson sampling interferometer design} \label{sec::shallow}

In Section~\ref{sec::extensions} we showed how to minimize the number of MZIs for boson sampling, and how to design universal interferometers by coupling smaller interferometers. Here we combine these to explore circuits that can perform boson sampling with several coupled interferometers.
\\

We make the following two assumptions. First, we assume that we have access to universal interferometers acting on $k$ modes for some integer $2 \leq k \leq m$. Based on the QR decomposition we know that they can transform any $k \times k$ matrix into a triangular one. Secondly, we are free to permute arbitrarily the modes when coupling interferometers together. \newline

Being able to perform unitary transformations on arbitrary sets of modes make it possible to design shallower circuits for boson sampling.  Given a rectangular matrix $V \in \mathbb{C}^{m \times n}$ representing a boson sampling experiment on $m$ modes with $n$ photons, the following greedy algorithm, inspired from \cite{de2021gaussian}, constructs a shallow linear optical circuit implementing $V$:
\begin{enumerate}
    \item assign to each row $i$ of $V$ a label $p_i$ such that $p_i = \min_j V[i,j] \neq 0$. If no such label exists then $p_i = n+1$.
    \item sort the labels in ascending order and permute the modes accordingly.
    \item define $i_{start} = \min(\min_i p_i = p_{i+1}, n-k)$
    \item starting from the mode $i_{start}$, group the modes by $k$. For one such group 
    \[ g_j = [i_{start}+jk, i_{start}+(j+1)k[ \] 
    look at the smallest label $q_j = \min_{i \in g_j} p_i$. If $q_j \neq n+1$ consider the submatrix 
    \[ W_j = V[g_j, q_j:\min(q_j+k, n)]. \]
    \item based on the QR decomposition, compute a unitary $U_j$ such that $U_jW_j$ is triangular, synthesize $U_j$ and add it to the circuit on the corresponding modes. 
    \item repeat steps 1-5 until $V$ is diagonal.
\end{enumerate}

\begin{figure*}
    \centering
    \includegraphics[scale=0.8]{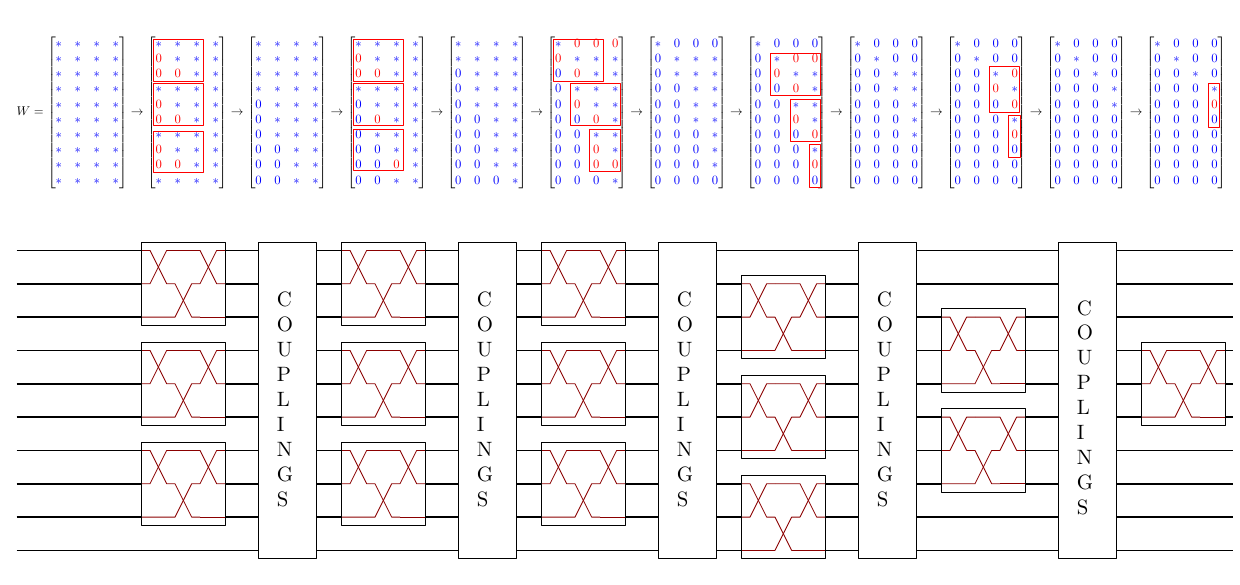}
    \caption{Example of boson sampling design with couplings for $m=10, n=4$ and universal interferometers on $3$ modes. Each interferometer triangularizes a subblock of the matrix to synthesize $W$. The couplings permute the modes. Once $W$ is triangular it is actually diagonal and the synthesis is done. Two extra couplings at the beginning and the end of the computation, not shown here, will be considered in our numerical analysis.}
    \label{fig::greedy}
\end{figure*}

Each iteration of the loop can only increase the number of zero elements in $V$. Once $V$ is upper triangular it will also be diagonal and the synthesis is complete. The inverse of the computed circuit gives a valid implementation of $V$. This process is shown in Fig.~\ref{fig::greedy}. \newline

The minimum MZI-depth is obtained in the case $k=2$, i.e, when we only couple single MZIs together. 

\begin{theorem}
\label{refthbs}
    Given layers of MZIs with arbitary connectivity, any non adaptive $m$-mode $n$-photon linear optical algorithm can be executed with a linear optical circuit of MZI-depth $O(n + \log(m))$.
\end{theorem}

\begin{proof}
    See Appendix~\ref{appendix}.
\end{proof}

This improves the $O(n\log(m))$ scheme proposed by Aaronson and Arkhipov \cite[Theorem 45]{aaronson2011computational} which, to our knowledge, is the shallowest analytical scheme in the literature with nonlocal MZIs. 

The work of \cite{go2024exploring} proposes a long range MZI architecture and provides numerical evidence that $O(\log(m))$ depth of this architecture suffices to mimic relevant properties of the Haar measure on $\mathcal{U}(m)$, and consequently  that boson sampling with average-case hardness could be performed at this depth with this architecture. In contrast,  Theorem \ref{refthbs} shows that we can do any $m$-mode transformation acting on $n$ photons in $O(n+\log(m))$ depth. Notably this means we can exactly sample outputs  of linear optical circuits whose unitaries are  chosen from the Haar measure over $\mathcal{U}(m)$.

Note that $O(mn)$ MZIs are necessary to implement an arbitrary $n$-photon $m$-mode experiment (see Section~\ref{sec::extensions}) and at most $m/2$ MZIs can be executed in one time-step. Hence, the minimum depth required to implement any unitary is $O(n)$. When $n=1$, the synthesis process essentially consists in zeroing $m-1$ coefficients of a vector. We can only half the number of nonzeros coefficients at each time step, resulting in a necessary $O(\log_2(m))$ depth. These two lower bounds give strong evidence that the depth complexity of Theorem \ref{refthbs} is the best achievable.

\section{Optimizing the total transmission in Boson sampling experiments} \label{sec::numerical}
In the following we perform a numerical analysis of coupled interferometer designs for boson sampling.  We show that with the scheme described in Section~\ref{sec::shallow} we can study the optimal chip-size for a given MZI loss and coupling loss. To reach a quantum computational advantage we likely need to perform boson sampling with $n \approx 50$ photons. Here we arbitrarily work with $n=48$. \newline

\paragraph*{\textbf{Total transmission.}} Our metric of interest is the total transmission of the interferometer. This relates directly to the probability of success of the experiment, where success means that no photon is lost. Let: 
\begin{itemize}
    \item $\eta_{mzi}$ be the transmission of one MZI,
    \item $\eta_{c}$ be the chip-to-chip coupling transmission, 
    \item $k$ the chip size,
    \item $d_{m,n}$ the depth, as an interferometer-based circuit, to perform the greedy algorithm on a matrix of size $m \times n$. We lack an analytical formula for $d_{m,n}$, so this is computed numerically. In the example of Figure~\ref{fig::greedy}, $d_{m,n} = 6$.
\end{itemize}

With extra in and out couplings, the total number of couplings is 
\[ n_{coupl} = d_{m, n} + 1. \]
The total MZI-depth is given by 
\[ d_{mzi} = d_{m,n} \times k. \]
The total transmission $\eta$ for one photon is given by 
\[ \eta = \eta_{mzi}^{d_{mzi}} \cdot \eta_{coupl}^{n_{coupl}} \]
and $\eta^n$ is the total transmission for $n$ photons. \newline 

\paragraph*{\textbf{Number of modes.}} In the usual formulation \cite{aaronson2011computational}, boson sampling is performed in the no-collision regime. With enough modes the probability that every photon ends up in a different mode can be made arbitrarily close to 1, removing the need for number resolving detectors. The necessary limit is $m = O(n^2)$ \cite{aaronson2011computational}, but practical results may be obtained with fewer modes \cite{bouland2023}. We study three cases:
\begin{itemize}
    \item $48$-photon $96$-mode ($m=2n$),
    \item $48$-photon $288$-mode ($m=6n$),
    \item $48$-photon $2304$-mode ($m=n^2$).
\end{itemize}

\paragraph*{\textbf{Target transmissions.}} 
With sufficient noise levels, boson sampling can be simulated efficiently with a classical computer. Loss reduces the maximum size of the matrices from which permanents have to be calculated \cite{renema2018classical}. For a transmission per photon $\eta < 0.85$, computing permanents of matrices no larger than $48$ is sufficient to simulate boson sampling. This may still be tractable with supercomputers. Therefore, to reach quantum utility, we require at least a total transmission per photon of $\eta = 0.85$. \newline

\paragraph*{\textbf{Numerics.}} 
We compute the following:

\begin{itemize}
    \item For fixed values of $\eta_{mzi}$ and $\eta_{c}$, we compute the chip size $k$ that maximizes the per-photon transmission.
    \item For a fixed value of $\eta_{mzi}$, we compute $\eta_{c}$ such that the per-photon transmission is $\eta = 0.85$, if such a value exists.
\end{itemize}

\begin{figure*}
\vspace*{-2cm}

\subfloat[96 modes. \label{heatmap1}]{\includegraphics[scale=0.6]{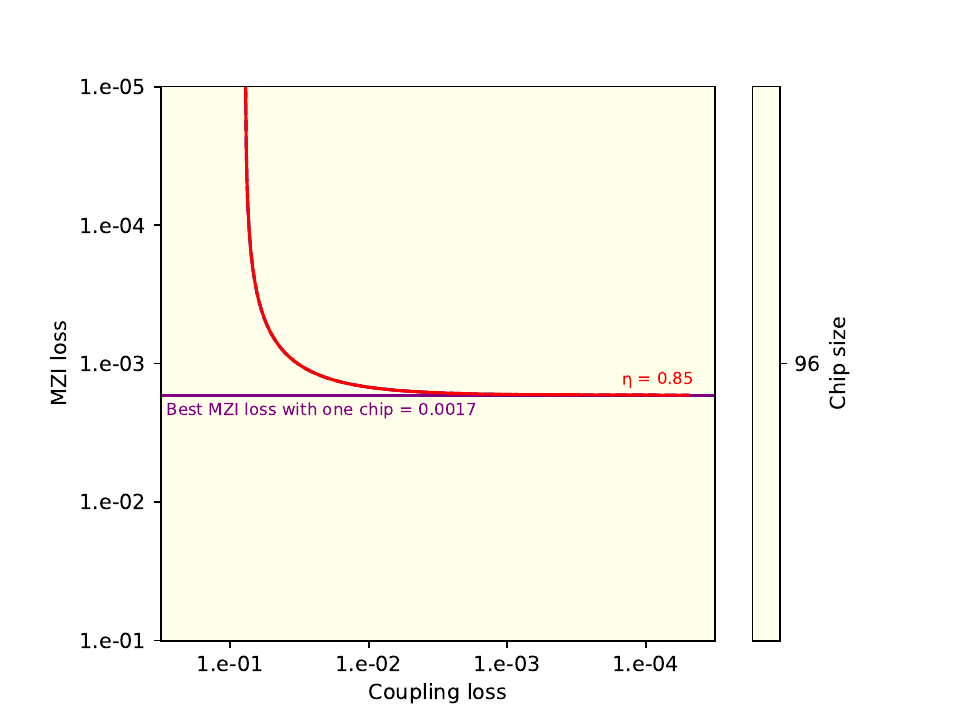}}

\subfloat[288 modes. \label{heatmap1}]{\includegraphics[scale=0.6]{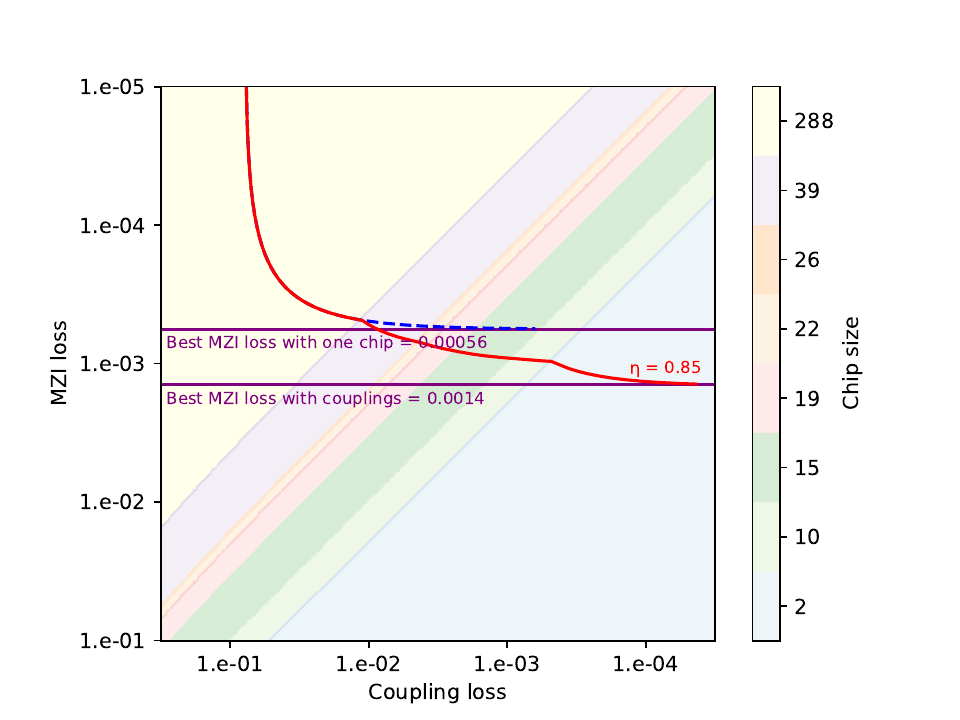}} 

\subfloat[2304 modes. \label{heatmap1}]{\includegraphics[scale=0.6]{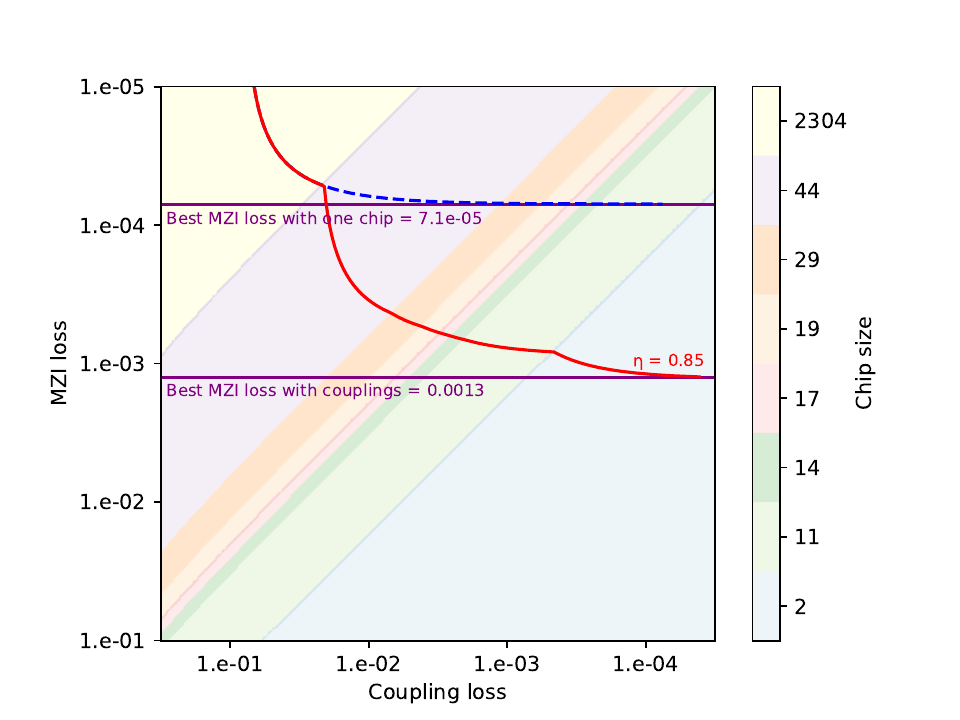}} 

\caption{$48$-photon boson sampling experiments. The heatmap in the background gives the block size that maximizes the transmission. In red the lines of constant transmission. In dashed blue the lines of constant transmission that correspond to always using one single chip to perform the experiment. The purple lines highlight the values of the MZI loss above which the target transmissions cannot be achieved. These plots account for the input-output coupling in the case of a single interferometer (a). }
\label{fig::heatmap}
\end{figure*}

These results are shown in the three graphs of Fig~\ref{fig::heatmap}: (a), (b), (c) correspond respectively to 96, 288 and 2304 modes. \newline 

The heatmap in the background gives the value of the chip size that maximizes the transmission. With $96$ modes, our scheme does not provide enough MZI-depth reduction and using one big interferometer maximises transmission. For $288$ and $2304$ modes there are cases where our scheme provides improvements when the coupling loss decreases but the MZI loss remains high. With the current values of transmissions ($97.6\%$ for MZIs and $86.5\%$ for the coupling \cite{maring2024versatile}), the optimal chip size for doing performing boson sampling on $2304$ modes would be $k=19$. 
The regime in which a single non-coupled interferometer maximises transmission shrinks with the number of modes, such that for boson sampling in the no-collision regime it is likely that a small chip size will be optimal. \newline

The red lines in Fig~\ref{fig::heatmap} correspond to the points of the plane of equal total transmission $\eta=0.85$. We observe that they correspond to piecewise exponential functions where the switches happen when the value of the optimal chip size changes. The dashed blue lines are the constant transmission lines for a single non-coupled interferometer. Provided that the coupling loss is small enough, our scheme reduces the hardware requirements on the MZI efficiency, up to an order of magnitude in the case of $2304$ modes. To reach $0.85$ transmission efficiency in a $48$-photon $2304$-mode experiment, if the coupling loss is sufficiently small (for instance around $10^{-3}$) then we can use MZIs that are $10$ times lossier than in a single $2304$-mode chip. 
\newline

More generally we can only trade MZI efficiency for coupling efficiency up to a point. We observe the two asymptotic behaviors when either the MZI or coupling loss tends to $0$. In both cases, there is an intrinsic limit where the decrease of one loss cannot compensate the increase of the other loss. In Fig~\ref{fig::heatmap} we have highlighted the three values of the MZI loss above which the target transmissions cannot be achieved. \newline

Overall, these results highlight the fact that to build an efficient Boson sampler, improving the transmission of individual MZIs is essential, but can be augmented with the development of low-loss fibre-chip coupling.

\section{Conclusion} \label{sec::conclu}

We proposed an alternative framework for the design of linear optical circuits. Our framework, based on recent results on CNOT circuits synthesis, consists of sorting labels in a network of conditional swaps while satisfying nearest neighbor architectural constraints. We managed to recover the best results of the literature in a simple framework that exhibits interesting theoretical properties: optimality in depth and count for MZI-based circuit, extension to partial synthesis for boson sampling, generalization to the coupling of multiple chips for designing larger circuits. \newline

We also proposed a new design for boson sampling experiments based on coupling multiple smaller interferometers. Whilst our scheme does not remove the need for advances in component losses, it enables some trade-off between MZI efficiency and coupling efficiency. 
\newline

As future work, it would be interesting to test other extensions of this framework. We think notably of the case of beam-splitter based linear optical circuit instead of MZI \cite{fldzhyan2020optimal}. It would be also interesting to extend the framework to take into account other kind of connectivities between the modes. 

\section*{Acknowledgments} The authors thank Boris Bourdoncle for insightful discussions and comments and Andreas Fyrillas and Simone Piacentini for their help on the hardware side. This work has been financed by the French government as part of France 2030 in the framework of UFOQO project.

\bibliography{Bruhat}

\onecolumngrid
\appendix

\section{Proof of the depth complexity of Boson sampling with long-range MZIs} \label{appendix}

Let $V \in \mathbb{C}^{m \times n}$ be our target isometry. In the special case where we couple MZIs, the greedy algorithm can be simplified as follows: 
\begin{itemize}
    \item assign to each row $i$ of $V$ the label 
    \[ p_i = \min_j V[i,j] \neq 0. \]
    If no such label exists then $p_i = n+1$.
    
    \item group rows with same labels two by two. For simplification, if row $i<n$ has label $p_i > i$ then we fix $p_i = i$. This will ensure us in the end that the nonzero part of the matrix is on the first $n$ rows. \\
    
    \item Let $i,j, i < j$ be such two rows with label $p_i$. Apply an long-range MZI between modes $i$ and $j$ to zero the entry $V[j, p_i]$.
    
    \item repeat steps 1-3 until $V$ is diagonal.
\end{itemize}

Overall, the whole process can be summarized by a sequence of integer arrays $(T_k)_k$, each $T_k$ being of size $n+1$, where $T_k[i]$ gives the number of rows with label $i$ at step $k$ of the process. \\

\begin{figure*}
    \centering
    \includegraphics[width=\linewidth]{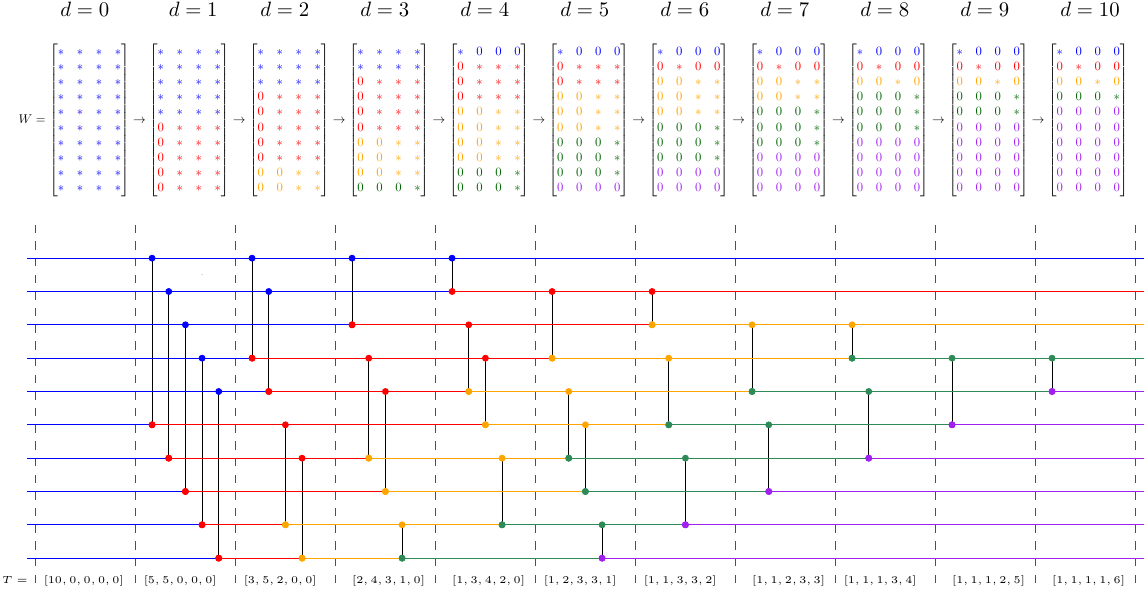}
    \caption{Application of the Boson sampling greedy framework for the special case of coupled MZIs. Alternatively, one can see the design as a synthesis problem over long-range MZIs. Each color corresponds to a different row label.}
    \label{fig:ccd}
\end{figure*}

A detailed example with $m=10, n=4$ is represented in Figure~\ref{fig:ccd} with values of the $T_k$ given.\\

Initially
\[ T_0 = [m, 0, \hdots, 0, 0] \] 
and the sequence obeys the following recursive formula:
\begin{align} 
T_{k+1}[i] = \begin{cases}
\ceil*{\frac{T_k[0]}{2}} & \text{ if } i = 0, \\ \\
\ceil*{\frac{T_k[i]}{2}}+ \floor*{\frac{T_k[i-1]}{2}} & \text{ if } 0 < i < n, \\ \\
T_k[i] + \floor*{\frac{T_k[i-1]}{2}} & \text{ if } i = n.
\end{cases} 
\label{recursive_T}
\end{align}

For some $K$ the sequence will reach the equilibrium

\[ T_K = [1,1,\hdots, 1, m-n] \]

and the goal of this section is to give a good estimation of $K$ as a function of $m$ and $n$. We look for a sufficiently tight upper bound. \\

We introduce the sequence of arrays $(S_k)$ where $S_k[i]$ is the parity of $T_k[i]$. We show by induction on $k$ a formula relating $T_k$ to the $S_l, l \leq k$.

\begin{lemma}
Let $k$ be any integer and $i < n$, 
\begin{equation} T_k[i] = \frac{1}{2^k} \binom{k}{i} m + \sum_{l=1}^k \frac{1}{2^l} \cdot \left( \sum_{j=0}^{\min(l,i)} \left[ \binom{l-1}{j} - \binom{l-1}{j-1} \right] S_{k-l}[i-j] \right). 
\label{eq1}
\end{equation}

\end{lemma}

\begin{proof}
Note that this formula does not work on the last entry of the $T_k$. We also assume that $\binom{N}{-1} = 0$ for some $N$. The formula is obviously true for $k = 0$. Now suppose this is true for some $k$. When $i=0$, the formula simplifies to 

\[ T_k[0] = \frac{1}{2^k} m + \sum_{l=1}^k \frac{1}{2^l} S_{k-l}[0]. \]

Using the fact that 

\[ \ceil*{\frac{T_k[i]}{2}} = \frac{T_k[i] + S_k[i]}{2} \]

it is easy to check that the formula is true for $k+1$ when $i=0$.

Now, let $0 < i < n$:

\begin{align*} 
T_{k+1}[i] & = \ceil*{\frac{T_k[i]}{2}}+ \floor*{\frac{T_k[i-1]}{2}} \\
& = \frac{T_k[i] + S_k[i] + T_k[i-1] - S_k[i-1]}{2} \\
& = \frac{1}{2} \left( \frac{1}{2^k} \binom{k}{i} m + \sum_{l=1}^k \frac{1}{2^l} \cdot \sum_{j=0}^{\min(l,i)} \left[ \binom{l-1}{j} - \binom{l-1}{j-1} \right] S_{k-l}[i-j] \right) \\ & + \frac{1}{2}\left(\frac{1}{2^k} \binom{k}{i-1} m + \sum_{l=1}^k \frac{1}{2^l} \cdot \sum_{j=0}^{\min(l,i-1)} \left[ \binom{l-1}{j} - \binom{l-1}{j-1} \right] S_{k-l}[i-1-j] \right) \\ & + \frac{S_k[i] - S_k[i-1]}{2}
\end{align*}

We merge two terms using Pascal's rule. We split the sums to deal the cases where $l < i$ and $l > i$ and we merge the two sums with a change of variable in the second sum, this gives 

\begin{align*} T_{k+1}[i] & = \frac{1}{2^{k+1}} \binom{k+1}{i} m + \frac{S_k[i] - S_k[i-1]}{2} \\ & +  \sum_{l=1}^{i-1} \frac{1}{2^{l+1}} \cdot \left( \sum_{j=0}^l \left[ \binom{l-1}{j} - \binom{l-1}{j-1} \right] S_{k-l}[i-j] + \sum_{j=1}^{l+1} \left[ \binom{l-1}{j-1} - \binom{l-1}{j-2} \right] S_{k-l}[i-j] \right) \\
& +  \sum_{l=i}^{k} \frac{1}{2^{l+1}} \cdot \left( \sum_{j=0}^i \left[ \binom{l-1}{j} - \binom{l-1}{j-1} \right] S_{k-l}[i-j] + \sum_{j=1}^{i} \left[ \binom{l-1}{j-1} - \binom{l-1}{j-2} \right] S_{k-l}[i-j] \right)
\end{align*}

Again, using Pascal's rule we can factor by $S_{k-l}[i-j]$. The cases $j=0$ in the two sums and the case $j=l+1$ in the first sum can be merged as well because of the nice behavior of the binomial coefficient. This gives
\begin{align*} T_{k+1}[i] & = \frac{1}{2^{k+1}} \binom{k+1}{i} m + \frac{S_k[i] - S_k[i-1]}{2} \\ & +  \sum_{l=1}^{i-1} \frac{1}{2^{l+1}} \cdot \left( \sum_{j=0}^{l+1} \left[ \binom{l}{j} - \binom{l}{j-1} \right] S_{k-l}[i-j] \right) \\
& +  \sum_{l=i}^{k} \frac{1}{2^{l+1}} \cdot \left( \sum_{j=0}^{i} \left[ \binom{l}{j} - \binom{l}{j-1} \right] S_{k-l}[i-j]\right) \\
& = \frac{1}{2^{k+1}} \binom{k+1}{i} m + \frac{S_k[i] - S_k[i-1]}{2} + \sum_{l=1}^k \frac{1}{2^{l+1}} \cdot \left( \sum_{j=0}^{\min(l+1,i)} \left[ \binom{l}{j} - \binom{l}{j-1} \right] S_{k-l}[i-j] \right)
\end{align*}

One last change of variable on $l$ and we can integrate the terms in $S_k$ in the sum to get the final result 

\begin{align*} T_{k+1}[i] & = \frac{1}{2^{k+1}} \binom{k+1}{i} m  + \frac{S_k[i] - S_k[i-1]}{2} + \sum_{l=2}^{k+1} \frac{1}{2^l} \cdot \left( \sum_{j=0}^{\min(l,i)} \left[ \binom{l-1}{j} - \binom{l-1}{j-1} \right] S_{k+1-l}[i-j] \right) \\
& = \frac{1}{2^{k+1}} \binom{k+1}{i} m + \sum_{l=1}^{k+1} \frac{1}{2^l} \cdot \left( \sum_{j=0}^{\min(l,i)} \left[ \binom{l-1}{j} - \binom{l-1}{j-1} \right] S_{k+1-l}[i-j] \right)
\end{align*}

and the formula is true for $k+1$.
\end{proof}

We can now prove a formula concerning the sum $\sum_{i=0}^{n-1} T_k[i]$.

\begin{lemma}
    Let $k$ be any integer, 

    \begin{equation} \sum_{i=0}^{n-1} T_k[i] = \sum_{i=0}^{n-1} \frac{1}{2^k} \binom{k}{i} \cdot m + \sum_{i=0}^{n-1} \sum_{l=i+1}^{k} \frac{1}{2^l}\cdot \binom{l-1}{i} \cdot S_{k-l}[n-i-1]. \end{equation}
\end{lemma}

\begin{proof}
    We only need to show 
    \[ \sum_{i=0}^{n-1} \sum_{l=1}^k \frac{1}{2^l} \cdot \left( \sum_{j=0}^{\min(l,i)} \left[ \binom{l-1}{j} - \binom{l-1}{j-1} \right] S_{k-l}[i-j] \right) = \sum_{i=0}^{n-1} \sum_{l=i+1}^{k} \frac{1}{2^l}\cdot \binom{l-1}{i} \cdot S_{k-l}[n-i-1]. \]

    Starting from the left-hand side, the proof essentially relies on a change of variable such that the most internal sum does not depend on the values of $S$ anymore. This gives

    \[ \sum_{i=0}^{n-1} \sum_{l=1}^k \frac{1}{2^l} \cdot S_{k-l}[i] \sum_{j=0}^{\min(l, n-i-1)} \left[ \binom{l-1}{j} - \binom{l-1}{j-1} \right] \]

    and the internal telescopic sum simplifies to 

    \[ \sum_{i=0}^{n-1} \sum_{l=1}^k \frac{1}{2^l} \cdot S_{k-l}[i] \cdot \binom{l-1}{\min(l, n-i-1)}. \]

    A final change of variable gives
    \[ \sum_{i=0}^{n-1} \sum_{l=1}^k \frac{1}{2^l} \cdot S_{k-l}[n-i-1] \cdot \binom{l-1}{\min(l, i)} \]

    and we get rid of the terms $l < i+1$ because the binomial coefficient is always zero. 
    
\end{proof}

Using Lemma 2 we derive an upper bound on $\sum_{i=0}^{n-1} T_k[i]$. Using the fact that $S_k[i] \leq 1$ we get 
\[ \sum_{i=0}^{n-1} T_k[i] \leq \sum_{i=0}^{n-1} \frac{1}{2^k} \binom{k}{i} \cdot m + \sum_{i=0}^{n-1}  \sum_{l=i+1}^k \frac{1}{2^l} \cdot \binom{l-1}{i}. \]

Using the fact that 
\[ \sum_{l=i+1}^k \frac{1}{2^l} \cdot \binom{l-1}{i} \leq \frac{1}{2} \sum_{l=0}^{\infty} \frac{1}{2^{l}} \cdot \binom{l}{i}  \]

we recognize the generating function of $\frac{y^i}{(1-y)^{i+1}}$ with $y=1/2$ such that we can finally conclude that 

\begin{equation} \sum_{i=0}^{n-1} T_k[i] \leq \sum_{i=0}^{n-1} \frac{1}{2^k} \binom{k}{i} \cdot m + n. \end{equation}

Given that the sequence $(\sum_{i=0}^{n-1} T_k[i])_k$ is a decreasing sequence of integers with limit $n$, it suffices to find an integer $K$ such that 

\begin{equation} \sum_{i=0}^{n-1} \frac{1}{2^K} \binom{K}{i} \cdot m < 1 \label{eq:ineq} \end{equation}

and we can conclude from 

\[ \sum_{i=0}^{n-1} T_K[i] < 1 + n \]

that $\sum_{i=0}^{n-1} T_K[i] = n$ and we indeed have reached the equilibrium. \\

To solve \[ \sum_{i=0}^{n-1} \frac{1}{2^K} \binom{K}{i} \cdot m < 1 \]

for $K$, we use the following inequality for the sum of binomial coefficients \cite{lovasz2003discrete} 

\[ \sum_{i=0}^{n-1} \binom{K}{i} \leq 2^{K-1} \exp\left(\frac{(K-2n)^2}{4(n-K)}\right). \]

Assuming $K > 2n$, we end up with a quadratic equation in $K$: 
\[ K^2 + \left(-4n - 4\ln(m/2)\right) K + 4n^2 + 4n\ln(m/2) > 0  \]

for which the smallest integer solution is 
\begin{equation} K = \ceil*{2n + 2\ln(m/2) + 2\sqrt{n\ln(m/2) + \ln(m/2)^2}}. \label{eq2} \end{equation}

\begin{figure}
    \centering
    \subfloat[]{\includegraphics[scale=0.6]{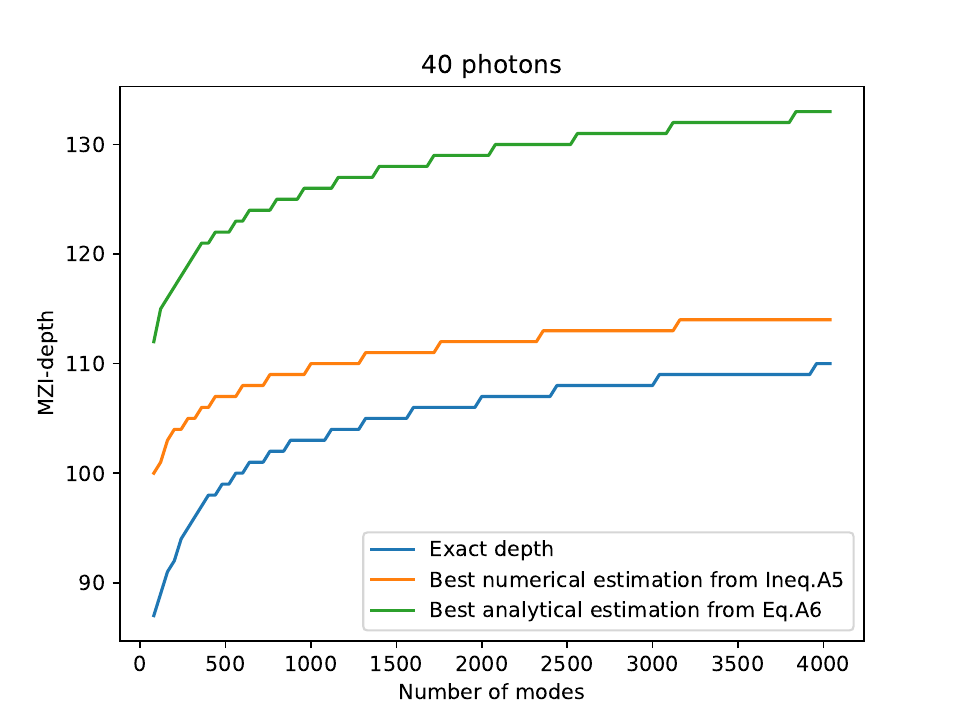}}
    \subfloat[]{\includegraphics[scale=0.6]{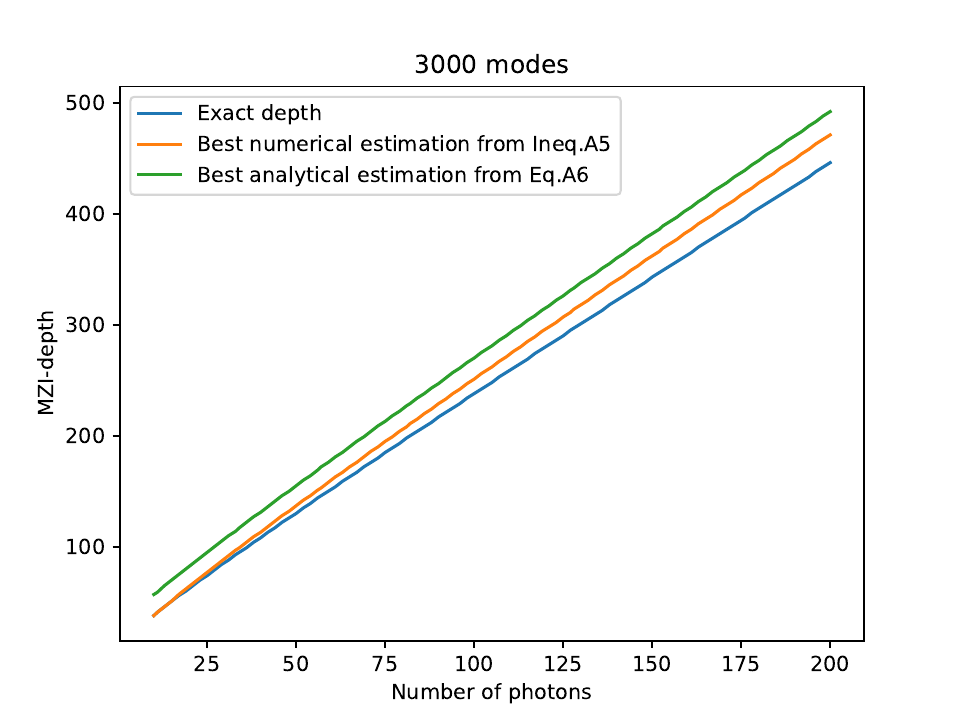}}
    \caption{Numerical evaluation of the tightness of the bounds of the MZI-depth in Boson sampling experiment with long-range MZIs.}
    \label{fig:tightness}
\end{figure}

We can compute numerically the exact depth, the best solution to inequality~\ref{eq:ineq} and the analytical estimation of Eq.~\ref{eq2}.
We plot these values for various values of $m$ in the case $n=40$ and for various values of $n$ for $m=3000$. They are represented in Figure~\ref{fig:tightness}. Overall the bounds follow a similar behavior and are quite tight.

\end{document}